\documentclass[12pt]{article}
\usepackage[top=2cm,bottom=2cm,left=1.5cm,right=1.5cm,columnsep=0.5cm]{geometry}
\usepackage{graphicx}
\usepackage{amsmath,amssymb}
\usepackage{bm}
\usepackage{hyperref}
\usepackage{booktabs}
\usepackage{xcolor}
\usepackage{microtype}
\usepackage{abstract}
\usepackage{tikz}
\usetikzlibrary{positioning,shapes.geometric}

\begin{document}

\title{Tensor Network Simulation of the Heisenberg Model\\
       on Heavy-Hex Lattices}

\author{
 Paolo D'Alberto\thanks{Advanced Micro Devices, Inc. (AMD).
 \texttt{paolo.dalberto@amd.com}.
 This work was developed in active collaboration with Claude
 (Anthropic)}\\
 \small Advanced Micro Devices, Inc.\\
 \small\texttt{paolo.dalberto@amd.com}
}

\date{\today}

\maketitle

\begin{abstract}
We extend CppSim, a high-performance C++/HIP tensor network simulator,
to the heavy-hex lattice geometry used in superconducting quantum processors.
The heavy-hex graph — a bipartite graph of degree-2 and degree-3 sites with
a natural 3-color gate schedule — is implemented as a drop-in grid class
with no changes to the gate application or belief propagation (BP) kernels.
We simulate the isotropic Heisenberg model on the \texttt{heavy\_hex\_3x3}
graph (68 sites, 76 bonds) in the Heisenberg picture using Pauli transfer
matrices (PTMs), and study the autocorrelation $C(t) = \langle Z_c(t)Z_c(0)\rangle$
and operator lightcone.
We incorporate a fully parametric hardware noise model: per-bond, per-color-class
$16\times 16$ PTMs are loaded from device characterization data and interleaved
with unitary gates, with noise scaling factor $\gamma \in \{1.0, 2.0, 3.0\}$
for multi-product formula (MPF) extrapolation to zero noise.
Chi convergence is demonstrated at $\chi=200$ ($|\mathrm{error}|<10^{-4}$,
reference $\chi=430$) on a 32\,GB GPU, with a $\sim$10$\times$ wall-clock
speedup over a Julia/AMDGPU.jl reference implementation at saturated bond dimension.
\end{abstract}

\section{Introduction}
\label{sec:intro}

Heavy-hex is the lattice topology used in superconducting quantum processors
\cite{heavyhex-topology}. Each qubit site has degree 2 or 3, and the graph admits
a natural 3-coloring of its bonds, enabling parallel gate application within
each color class.

Classical tensor network simulation of quantum circuits on heavy-hex graphs
serves two purposes: (i) validation of quantum computations for small system sizes
where exact simulation is feasible
\cite{tindall2023gauging,tindall2024eagle,tindall2025disordered,rudolph2025simulating},
and (ii) computation of noise-free reference values for quantum error mitigation
protocols such as the multi-product formula (MPF) \cite{mpf-ref}.

In this paper we show that CppSim \cite{dalberto2026cppsim}, originally developed for
rectangular 2D Ising dynamics, extends naturally to heavy-hex topology with
minimal code changes. The hot-path kernels — gate application, BP,
workspace management — are topology-agnostic and require no modification.
Only the grid construction and Hamiltonian gate matrix need to be defined
for the new geometry. We further incorporate a fully parametric hardware noise
model via Pauli transfer matrices, and demonstrate gamma-MPF extrapolation
to zero noise using three noise amplification levels $\gamma \in \{1, 2, 3\}$.

\section{Heavy-Hex Lattice and Hamiltonian}
\label{sec:grid}

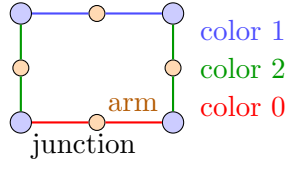
\begin{figure}[t]
\centering
\begin{tikzpicture}[scale=0.72,
    junction/.style={circle,draw,fill=blue!20,minimum size=8pt,inner sep=0pt},
    arm/.style={circle,draw,fill=orange!30,minimum size=6pt,inner sep=0pt},
    c0/.style={thick,red},
    c1/.style={thick,blue!70},
    c2/.style={thick,green!60!black}]
\def\dx{1.4}\def\dy{1.0}
\node[junction] (J00) at (0,0) {};
\node[junction] (J10) at (2*\dx,0) {};
\node[junction] (J01) at (0,2*\dy) {};
\node[junction] (J11) at (2*\dx,2*\dy) {};
\node[arm] (A0t) at (\dx,0) {};       
\node[arm] (A0b) at (\dx,2*\dy) {};   
\node[arm] (A0l) at (0,\dy) {};       
\node[arm] (A0r) at (2*\dx,\dy) {};   
\draw[c0] (J00) -- (A0t) -- (J10);
\draw[c1] (J01) -- (A0b) -- (J11);
\draw[c2] (J00) -- (A0l) -- (J01);
\draw[c2] (J10) -- (A0r) -- (J11);
\node[right] at (2*\dx+0.3,0.3) {\small\textcolor{red}{color 0}};
\node[right] at (2*\dx+0.3,2*\dy-0.3) {\small\textcolor{blue!70}{color 1}};
\node[right] at (2*\dx+0.3,\dy) {\small\textcolor{green!60!black}{color 2}};
\node[below right,font=\small] at (J00) {junction};
\node[above right,font=\small,orange!70!black] at (A0t) {arm};
\end{tikzpicture}
\caption{One unit cell of the heavy-hex graph. Junction sites (blue, degree 3)
connect via arm sites (orange, degree 2). Bonds are 3-edge-colorable (red, blue, green),
enabling parallel gate application within each color class.}
\label{fig:heavyhex}
\end{figure}

The \texttt{heavy\_hex\_$n_x \times n_y$} graph $G=(V,E)$
(Fig.~\ref{fig:heavyhex}) has
$n_x \cdot n_y$ \emph{junction} sites of degree 3 and
$(2n_xn_y + n_x + n_y)$ arm sites of degree 2,
for a total of $|V| = 3n_xn_y + n_x + n_y$ sites and
$|E| = 2n_xn_y + n_x + n_y$ bonds.
For $n_x = n_y = 3$: $|V|=68$ ($16$ junction $+ 52$ arm) and $|E|=76$.
We write $\partial v \subset E$ for the set of bonds incident to site $v$.
The graph is bipartite and 3-edge-colorable — bonds partition into 3 color
classes of sizes 25, 25, 26, corresponding to the natural gate layers
of the hardware schedule.

Site tensors carry the evolved Pauli operator in the Heisenberg picture,
with shapes $[\chi^2, D]$ (arm) and $[\chi^3, D]$ (junction),
where $D=4$ is the Pauli basis dimension $\{I, X, Y, Z\}$.
The \texttt{FlatHeavyHexGrid} class inherits from \texttt{Grid} and populates
the same \texttt{Site}/\texttt{Bond}/\texttt{Gate} data structures as the
rectangular grid, with no changes to the gate, belief propagation, or
workspace kernels.

We simulate the isotropic Heisenberg (XXX) model:
\begin{equation}
    H = J \sum_{\langle ij \rangle} \bigl(X_i X_j + Y_i Y_j + Z_i Z_j\bigr)
\end{equation}
with bond-specific coupling $J_{\langle ij\rangle}$ from device characterization
and Trotter step $\varepsilon$.
In the Heisenberg picture, the two-site gate
$U = e^{-i\varepsilon J(X\otimes X + Y\otimes Y + Z\otimes Z)}$
is a $4\times4$ unitary acting on the two-qubit Hilbert space $\mathbb{C}^2\otimes\mathbb{C}^2$.
To track operator dynamics we work in the two-site Pauli basis
$\{P_\mu \otimes P_\nu\}$ with $P_\mu \in \{I,X,Y,Z\}$,
which has $4\times4 = 16$ elements — hence the state vector at each bond has
dimension 16, and the gate acts as a $16\times16$ real Pauli transfer matrix (PTM)
via $O \mapsto U^\dagger O\, U$.
Here and below we use the physics shorthand $AB \equiv A\otimes B$,
so $IX \equiv I\otimes X$, $YZ \equiv Y\otimes Z$, etc.

The PTM is block-diagonal because the XXX interaction conserves the
\emph{Pauli commutation signature}: for any two-site operator $O$,
if $[H, O] = 0$ then $U^\dagger O\, U = O$ (invariant).
The four operators $\{II, XX, YY, ZZ\}$ all commute with $H$,
forming a $4\times4$ identity block.
The remaining 12 operators decompose into three invariant
\emph{mixing} subspaces of size 4, determined by which single-site
Pauli is ``odd'' under the symmetry.
For example, the commutators of $H$ with each of
$\{IX, XI, YZ, ZY\}$ produce linear combinations
within that same set, so it is closed under the adjoint action
and evolves as a $4\times4$ rotation matrix.
Cyclic permutation $X\!\to\!Y\!\to\!Z\!\to\!X$ yields the other two blocks
$\{IY,YI,XZ,ZX\}$ and $\{IZ,ZI,XY,YX\}$.
In total the PTM is block-diagonal with one $4\times4$ identity block
and three $4\times4$ rotation blocks, all real.

\section{Hardware Noise Model}
\label{sec:noise}

We extend the simulation to include hardware noise channels characterized on the
target device. In the Pauli transfer matrix (PTM) formalism, a noise channel $\mathcal{E}$
acting on a two-site operator is represented as a $16\times16$ real matrix
$E_{ij} = d^{-1}\,\mathrm{Tr}(P_i^\dagger\,\mathcal{E}(P_j))$,
where $\{P_i\}$ is the two-site Pauli basis.
Since both operators and noise channels are Hermitian-preserving, $E$ is real
throughout, consistent with our float32 simulation.

Each bond implements the two-qubit interaction through a hardware decomposition
$U_{\mathrm{bond}} = R_{xx}(\alpha)\cdot R_{yy}(\alpha)\cdot R_{zz}(\delta)$,
where $\alpha = -J_{\langle ij\rangle}\cdot\varepsilon$ scales with the bond coupling
and Trotter step $\varepsilon$, while $\delta = -J_{\langle ij\rangle}$ is
independent of $\varepsilon$.
Noise is interleaved after each sub-gate, and the full noisy gate for bond $b$
in its color class is:
\begin{equation}
    \mathrm{PTM}_{\mathrm{noisy}}^{(b)} =
    G_1 \cdot M_n^{(b)} \cdot G_2(\alpha_b) \cdot M_n^{(b)}
    \cdot G_3(\alpha_b) \cdot M_n^{(b)} \cdot G_4(\delta_b)
\end{equation}
where $M_n^{(b)}$ is the hardware noise PTM for bond $b$ in its color class,
and $G_1$--$G_4$ implement the hardware sub-layers.
After all color classes, a kick gate $R_x(2\phi)$ is applied once per site
($\phi = 1.41371669$).

The noise characterization is fully parametric: a configuration file specifies
an independent $16\times16$ PTM for each (bond, color class) pair and noise
scaling factor $\gamma \in \{1.0, 2.0, 3.0\}$ ($\gamma=1.0$: hardware noise
as measured; $\gamma=2.0, 3.0$: amplified for MPF extrapolation).
This single composed PTM is pre-computed per bond at build time,
so runtime applies exactly one matrix--vector product per bond per Trotter step —
the same cost as the noiseless simulation.

\section{Autocorrelation and Chi Convergence}
\label{sec:ct}

We initialize the Heisenberg-picture operator as $Z$ at a single center site
(site 13, a degree-3 junction at the graph midpoint) and $I$ elsewhere.
The autocorrelation $C(t) = \langle Z_c(t) Z_c(0) \rangle$ is computed
via the Bethe approximation \cite{tindall2023gauging,evenbly2026loop} after each Trotter step.
We study convergence with respect to the bond dimension $\chi$ by running
nine simulations with $\chi \in \{50, 100, 150, 200, 250, 300, 350, 400, 430\}$
over 30 Trotter steps at Trotter step $\varepsilon = 0.25$ and cutoff $10^{-4}$.

\begin{figure}[h]
    \centering
    \includegraphics[width=0.80\columnwidth]{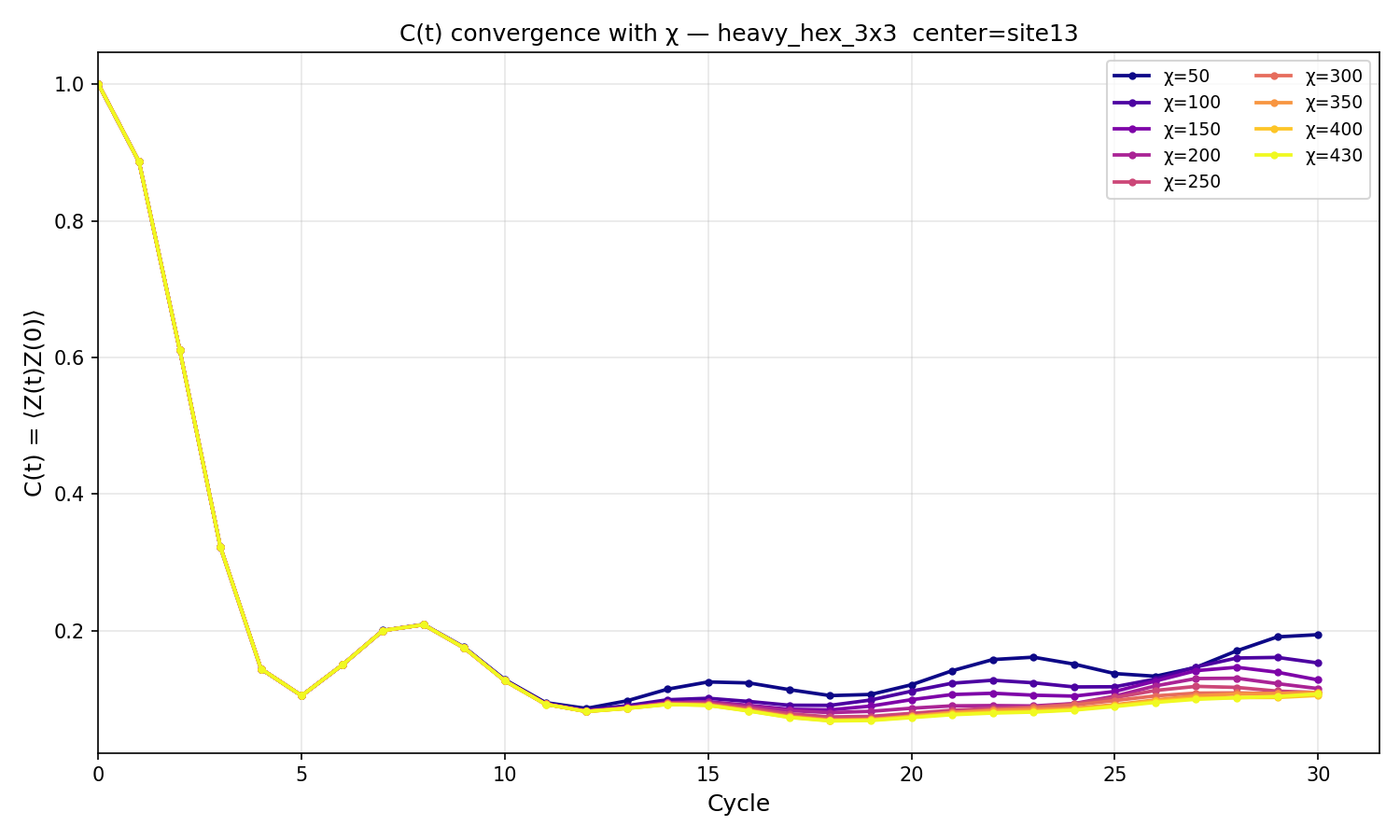}
    \caption{Autocorrelation $C(t) = \langle Z_c(t) Z_c(0) \rangle$ for
    $\chi = 50, 100, 150, 200, 250, 300, 350, 400, 430$ over 30 Trotter steps
    at $\varepsilon=0.25$.
    The oscillatory behaviour is a genuine quantum recurrence of the XXX model
    on the finite heavy-hex graph. Curves for $\chi \geq 150$ are visually
    indistinguishable, indicating convergence.}
    \label{fig:ct}
\end{figure}

Figure~\ref{fig:ct} shows $C(t)$ for $\chi = 50, 100, 150, 200, 250, 300,
350, 400, 430$ over 30 Trotter steps.
The result converges to $\lesssim 10^{-4}$ at $\chi = 150$ and to
$3\times10^{-6}$ at $\chi = 400$ (reference $\chi=430$),
indicating that $\chi_{\max} = 200$ is sufficient for
$|{\rm error}| < 10^{-4}$ on this graph at 10 cycles.

The non-monotonic behaviour of $C(t)$ — a decrease followed by partial
recovery — is a genuine quantum recurrence of the XXX model on the degree-3
lattice, confirmed by its independence from $\chi$.

To put the MPF results of Section~\ref{sec:mpf} in perspective: the
gamma-MPF simulations use $\varepsilon=0.1$, so 10 Trotter cycles
correspond to physical time $T = 10 \times 0.1 = 1.0$, which maps to
only 4 steps on the axis of Figure~\ref{fig:ct} (where $\varepsilon=0.25$).
The MPF operates entirely within the initial decay regime, well before
the quantum recurrence sets in around step 10.

\begin{figure}[t]
    \centering
    \includegraphics[width=\textwidth]{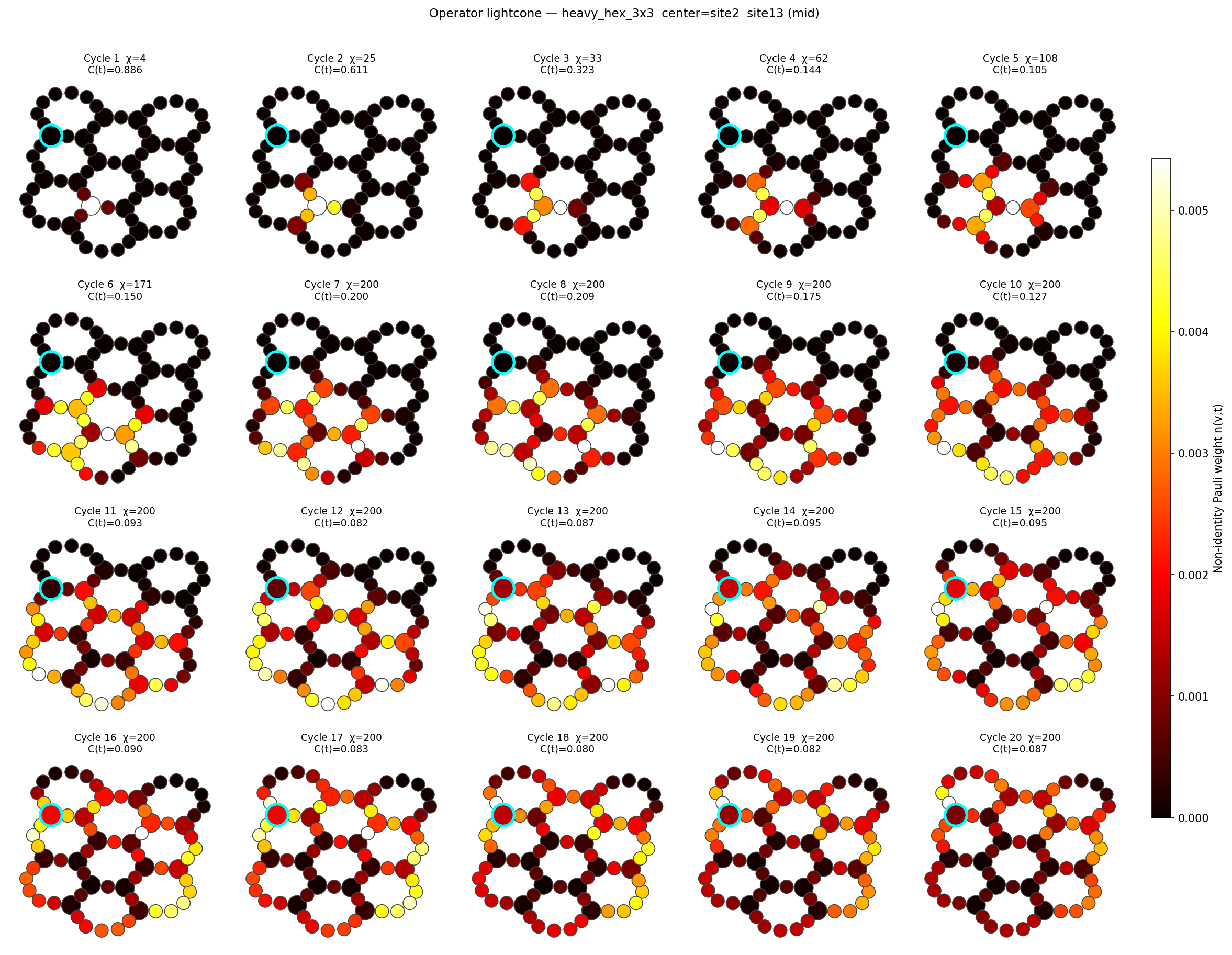}
    \caption{Non-identity Pauli weight $n(v,t)$ at each site of the
    heavy\_hex\_3x3 graph over 20 Trotter steps, initialised at site 13
    (mid-graph degree-3 junction). The operator spreads radially
    outward with 3-fold hexagonal symmetry, reaching the graph boundary
    by cycle $\sim$17. Color scale is per-panel normalized.}
    \label{fig:lightcone}
\end{figure}

Figure~\ref{fig:lightcone} shows the non-identity Pauli weight $n(v,t)$
at each site for 20 cycles. The operator spreads radially outward
from the center junction, following the 3-fold symmetry of the hexagonal
lattice. The wavefront reaches the boundary sites by cycle 17,
defining the computational window within which the simulation is exact
regardless of $\chi$ truncation at the boundary.

\section{Performance}
\label{sec:perf}

We report wall-clock times on a 32\,GB GPU using Householder QR decomposition (QR) and float32 arithmetic.
All timings are per Trotter cycle at saturated bond dimension.

\begin{table}[h]
\centering
\small
\caption{Wall-clock time per cycle. GPU runs: $\chi$ capped at 430,
float32, Householder QR, cutoff=$10^{-4}$.
CPU: float64, no $\chi$ cap (grows freely to $\chi_{\rm CPU}$).
Julia: GPU, noiseless evolution time only.
`---': not measured.}
\label{tab:comparison}
\begin{tabular}{rr rrr rr}
\toprule
Cycle & $\chi_{\rm max}$ & CPU (s) & $\chi_{\rm CPU}$ & Julia (s) & CppSim (s) & Speedup \\
\midrule
 2 &  14 &      9 &  14 &  29 &   0.1 & 290$\times$ \\
 4 &  60 &     15 &  58 &  29 &   0.5 &  58$\times$ \\
 6 & 165 &    201 & 202 &  38 &   1.9 &  20$\times$ \\
 7 & 278 &    978 & 364 &  65 &   4.4 &  15$\times$ \\
 8 & 430 &  4{,}393 & 637 & 132 & 12.3 &  11$\times$ \\
 9 & 430 &    --- & --- & 251 &  25.2 &  10$\times$ \\
10 & 430 &    --- & --- & 418 &  46.9 &   9$\times$ \\
\bottomrule
\end{tabular}
\end{table}

All runs saturate at their $\chi$ cap by cycle 7--9.  The dominant
cost is the Householder QR decomposition ($\mathcal{O}(m n^2)$ where
$m = \chi^2 D$ for degree-3 sites), scaling approximately as $\chi^3$.
Table~\ref{tab:comparison} compares wall-clock times at $\chi=430$,
cutoff=$10^{-4}$. Julia times report autocorrelation evolution only.
At saturated bond dimension (cycles 8--10), CppSim achieves a
consistent $\sim$10$\times$ speedup. At saturated bond dimension
($\chi = 430$, cycles 8--10), CppSim achieves a $\sim$10$\times$
speedup over the Julia/AMDGPU.jl reference implementation on the same
GPU. Both use Householder QR and float32 arithmetic. The speedup
arises from: (i) zero GPU memory allocation during simulation
(pre-allocated workspace), (ii) direct HIP/rocBLAS calls without
framework dispatch overhead, and (iii) on-device \texttt{pseudo\_sqrt}
kernel eliminating CPU--GPU transfers.

\newpage
\section{Multi-Product Formula Application}
\label{sec:mpf}

The gamma-MPF uses three noisy states at noise scaling factors
$\gamma \in \{1.0, 2.0, 3.0\}$ with fixed Trotter step $\varepsilon=0.1$,
extrapolating to $\gamma=0$ to recover the noise-free signal.
The MPF extrapolation requires coefficients $\{c_i\}$ such that
$C_{\mathrm{mpf}}(t) = \sum_i c_i C_n(\gamma_i, t) \approx C_{\mathrm{nl}}(t)$,
with constraint $\sum_i c_i = 1$.
\begin{figure}[ht]
    \centering
    \includegraphics[width=\columnwidth]{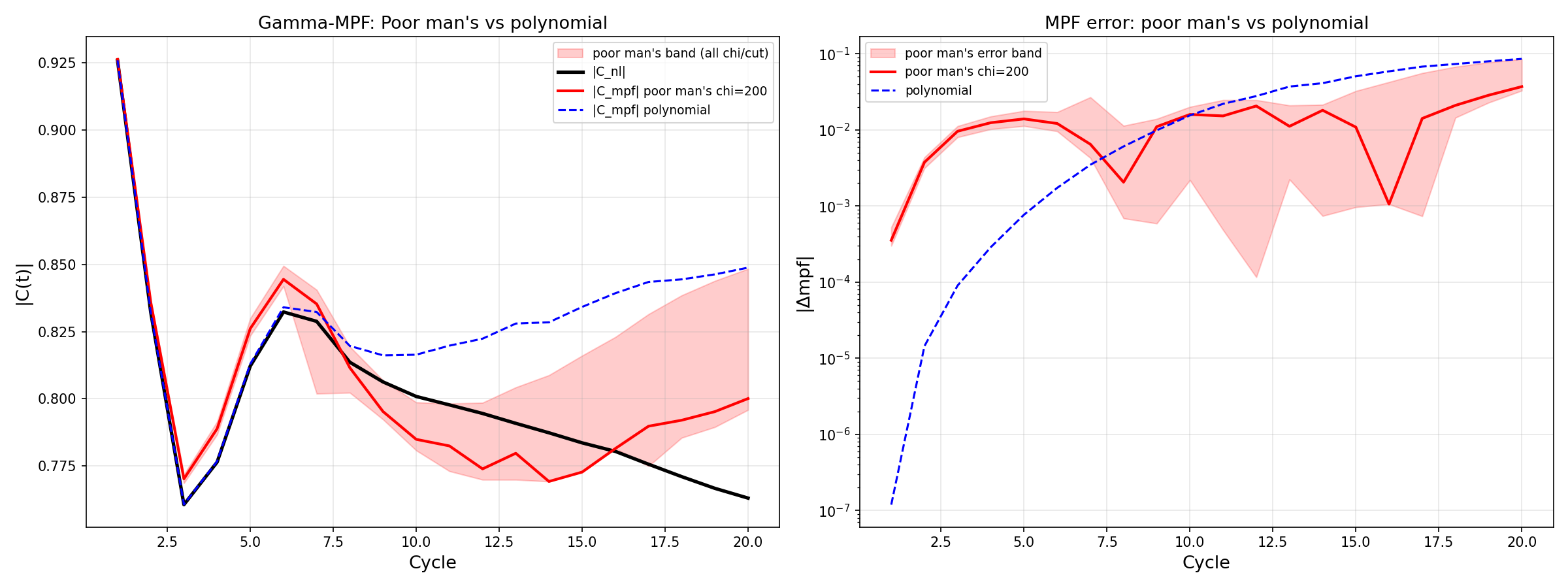}
    \caption{MPF error $|C_{\mathrm{mpf}}(t) - C_{\mathrm{nl}}(t)|$ (log scale)
    for gamma-MPF with $\chi=200$, $\varepsilon=0.1$, 20 Trotter cycles.
    Left: $|C_{\mathrm{mpf}}(t)|$ vs $|C_{\mathrm{nl}}|$ reference (dashed).
    Right: error comparison — poor man's MPF (red, correlation matrix)
    outperforms polynomial extrapolation (blue) by 2--3 orders of magnitude
    up to cycle 12. Band shows variation across all $\chi$/cutoff values.}
    \label{fig:mpf}
\end{figure}

Rather than computing quantum inner products $\langle\psi_i|\psi_j\rangle$
(which requires solving a gauge-dependent cross-state contraction problem),
we construct the coefficient matrix directly from the observable time series.
We define the correlation matrix:
\begin{equation}
    A_{ij} = \sum_t C_n(\gamma_i, t)\, C_n(\gamma_j, t), \qquad
    B_i    = \sum_t C_n(\gamma_i, t)\, C_{\mathrm{nl}}(t)
\end{equation}
and solve $\min_{\mathbf{c}} \|A\mathbf{c} - \mathbf{B}\|^2$ subject to
$\sum_i c_i = 1$.
This \emph{poor man's MPF} is gauge-invariant — it depends only on the
autocorrelation $C(t)$, a physical observable — and requires no additional
computation beyond the Trotter evolution already performed.

This formulation is also a natural scaffold for the full quantum inner product
MPF: replacing $A_{ij}$ with $\langle\psi_i|\psi_j\rangle$ recovers the
complete formulation used in \cite{mpf-ref}.
Below we describe the algorithm and block structure for computing these
inner products efficiently on GPU.

\subsection*{Quantum inner product via double-layer belief propagation}

The inner product between two Heisenberg-picture operator states
$|\psi_a\rangle$ and $|\psi_b\rangle$ — both living on the heavy-hex graph
as tensor networks with bond dimension $\chi$ — is computed by contracting
a \emph{double-layer} tensor network.
At each site $v$, the single-layer site tensor $\psi_v[\mathbf{i}, d]$
(bond indices $\mathbf{i}$, Pauli index $d\in\{I,X,Y,Z\}$) becomes:
\begin{equation}
    T_v[\mathbf{i}_a, \mathbf{i}_b]
    \;=\; \sum_{d=1}^{D} \psi_a[v, \mathbf{i}_a, d]\;\psi_b[v, \mathbf{i}_b, d]
    \label{eq:Tv}
\end{equation}
where $\mathbf{i}_a$ and $\mathbf{i}_b$ are \emph{independent} bond indices
for the two layers: $i_{a,k} \in [0, \chi_{a,k})$ is determined by the bond
dimension of $|\psi_a\rangle$, and $i_{b,k} \in [0, \chi_{b,k})$ by that of
$|\psi_b\rangle$. These may differ — for cross-overlaps between the noiseless
and noisy states, $\chi_{a,k} \neq \chi_{b,k}$ in general.
Each bond $e\in E$ carries a $\chi_{a,e}\times\chi_{b,e}$ message matrix
$M_e[i_a,i_b]$, and the inner product follows from the Bethe partition function:
\begin{equation}
    \langle\psi_a|\psi_b\rangle
    \;\approx\;
    \frac{\prod_{v\in V} a_v}{\prod_{e\in E} z_e},
    \label{eq:bethe_formula}
    \qquad
    a_v = \sum_{\mathbf{i}_a,\mathbf{i}_b} T_v\!\prod_{e\in\partial v} M_e,
    \quad
    z_e = \sum_{i_a,i_b} M_e^{\to}[i_a,i_b]\,M_e^{\leftarrow}[i_a,i_b]
\end{equation}
which is the BP approximation (exact on trees, approximate on the loopy graph $G$).

\paragraph{Message update without building $T_v$.}
Each bond $e$ carries \emph{two} directed messages: $M_e^{\to}[i_a,i_b]$
flowing from site $a$ to site $b$, and $M_e^{\leftarrow}[i_a,i_b]$ in the
reverse direction.
Both are $\chi_{a,e}\times\chi_{b,e}$ matrices coupling the bond index
of $|\psi_a\rangle$ to that of $|\psi_b\rangle$.

\emph{Initialization.}
Before any BP sweep, the messages are warm-started from the single-layer
BP messages already computed during the Trotter step:
\begin{equation}
    M_e^{(0)}[i_a,\, i_b]
    \;=\;
    \bigl|m_e^{(a)}[i_a]\bigr| \cdot \bigl|m_e^{(b)}[i_b]\bigr|
    \label{eq:msg_init}
\end{equation}
where $m_e^{(a)}$ and $m_e^{(b)}$ are the diagonal entries of the converged
single-layer BP messages for $|\psi_a\rangle$ and $|\psi_b\rangle$ respectively.
This rank-1 warm-start — a Kronecker product of the two single-layer diagonals
— places BP in a good basin of attraction at zero additional cost.
After convergence, $M_e$ evolves to a general matrix of rank up to
$\min(\chi_{a,e}, \chi_{b,e}, D)$.

\emph{Update rule.}
Constructing $T_v$ explicitly costs $O(\chi^{2\,\text{deg}})$ — for degree-3
junction sites this is $O(\chi^6)$, which is intractable at $\chi\gtrsim 10$.
Instead, we contract $M$ into $\psi$ \emph{before} forming the outer product,
giving $O(\chi^3 D)$ per message update.

Consider a degree-2 (arm) site $v$ with bonds $k_0$ and $k_1$.
To update the outgoing message on bond $k_0$, we use only the
\emph{incoming} message from the other bond $k_1$ — BP never uses the
current message of a bond to update itself.
The update proceeds in two steps:
\begin{align}
    \phi_a[i_{a,0},\, j_{b,1},\, d]
    &= \sum_{i_{a,1}} \psi_a[v,\, i_{a,0},\, i_{a,1},\, d]
       \cdot M_{k_1}[i_{a,1},\, j_{b,1}]
    \label{eq:step1}\\
    M_{k_0}^{\rm new}[i_{a,0},\, i_{b,0}]
    &= \sum_{j_{b,1},\, d} \phi_a[i_{a,0},\, j_{b,1},\, d]
       \cdot \psi_b[v,\, i_{b,0},\, j_{b,1},\, d]
    \label{eq:step2}
\end{align}
Here $i_{a,k}$ denotes the bond-$k$ index of state $|\psi_a\rangle$
and $j_{b,k}$ that of state $|\psi_b\rangle$
(e.g.\ $i_{a,0}\in[0,\chi_{a,0})$ for bond $k_0$ in $|\psi_a\rangle$,
$j_{b,1}\in[0,\chi_{b,1})$ for bond $k_1$ in $|\psi_b\rangle$).
Equation~(\ref{eq:step1}) absorbs the incoming message $M_{k_1}$ into
$\psi_a$ by summing over $i_{a,1}$, producing an
intermediate tensor $\phi_a$ of shape $[\chi_{a,0}, \chi_{b,1}, D]$.
Equation~(\ref{eq:step2}) pairs $\phi_a$ with $\psi_b$ by summing over
the shared index $j_{b,1}$ and the Pauli index $d$, yielding the new
outgoing message $M_{k_0}^{\rm new}$ of shape $[\chi_{a,0}, \chi_{b,0}]$.
The symmetric update for $M_{k_1}^{\rm new}$ follows by swapping $k_0\leftrightarrow k_1$.
For degree-3 (junction) sites, two sequential contractions of the form
(\ref{eq:step1}) absorb the two incoming messages before the final
step~(\ref{eq:step2}), keeping the cost at $O(\chi^3 D)$ throughout.

\paragraph{Matrix formulation.}
Both the message update and the Bethe PF reduce to standard matrix
multiplications.
We use the shorthand $\tilde{\Psi}_a \in \mathbb{R}^{\chi_{a,0} \times \chi_{a,1}D}$
for the site tensor $\psi_a[v, i_{a,0}, i_{a,1}, d]$ reshaped by treating
the last bond index and Pauli index $d\in\{I,X,Y,Z\}$ jointly.

\textit{Degree-2 site} (arm node, bonds $k_0$ and $k_1$; updating $M_{k_0}^{\rm new}$
using incoming $M_{k_1}$):
\begin{align}
    \Phi_a &= \tilde{\Psi}_a \cdot (M_{k_1} \otimes I_D)
    \;\in\; \mathbb{R}^{\chi_{a,0} \times \chi_{b,1}D}
    \quad\text{(GEMM, batched over }D\text{)}
    \label{eq:gemm1}\\
    M_{k_0}^{\rm new} &= \Phi_a \cdot \tilde{\Psi}_b^{\top}
    \;\in\; \mathbb{R}^{\chi_{a,0} \times \chi_{b,0}}
    \quad\text{($D$ sequential GEMMs)}
    \label{eq:gemm2}
\end{align}

\textit{Degree-3 site} (junction, bonds $k_0$, $k_1$, $k_2$; updating $M_{k_0}^{\rm new}$
using incoming $M_{k_1}$ and $M_{k_2}$):
same structure applied twice before the final step:
\begin{align}
    \Phi_a^{(1)} &= \tilde{\Psi}_a^{[k_2]} \cdot (M_{k_2}\otimes I_D)
    \;\in\; \mathbb{R}^{\chi_{a,0}\chi_{a,1}\times\chi_{b,2}D}
    \label{eq:gemm3}\\
    \Phi_a^{(2)} &= \Phi_a^{(1)[k_1]} \cdot (M_{k_1}\otimes I_D)
    \;\in\; \mathbb{R}^{\chi_{a,0}\times\chi_{b,1}\chi_{b,2}D}
    \label{eq:gemm4}\\
    M_{k_0}^{\rm new} &= \Phi_a^{(2)} \cdot \tilde{\Psi}_b^{\top}
    \;\in\; \mathbb{R}^{\chi_{a,0}\times\chi_{b,0}}
    \label{eq:gemm5}
\end{align}
where $[\,\cdot\,]$ denotes a reshape placing that bond index outermost.
Total cost: $O(\chi^3 D)$ for both degree-2 and degree-3.
The formulation above was derived by formalising the implemented and
validated GPU code (\texttt{test\_blf\_sgemm.cpp}); the equations
describe exactly what the kernels compute.

\textit{Bethe partition function.}
The vertex and edge factors in~(\ref{eq:bethe_formula}) are computed
by the same GEMM chain applied to \emph{all} bonds:
\begin{align}
    a_v &= \langle M_{k_0},\;
           \Phi_a^{\rm full} \cdot (\tilde{\Psi}_b^{\rm full})^{\top}
           \rangle_F
    \label{eq:bethe_v}\\
    z_e &= \langle M_e^{\to},\; M_e^{\leftarrow}\rangle_F
    \label{eq:bethe_e}
\end{align}
where $\langle\cdot,\cdot\rangle_F$ is the Frobenius inner product
(one \texttt{sdot} call per bond).
The Bethe PF costs the same as one message update per site — $O(\chi^3 D)$.

\begin{figure}[h]
    \centering
    \includegraphics[width=\columnwidth]{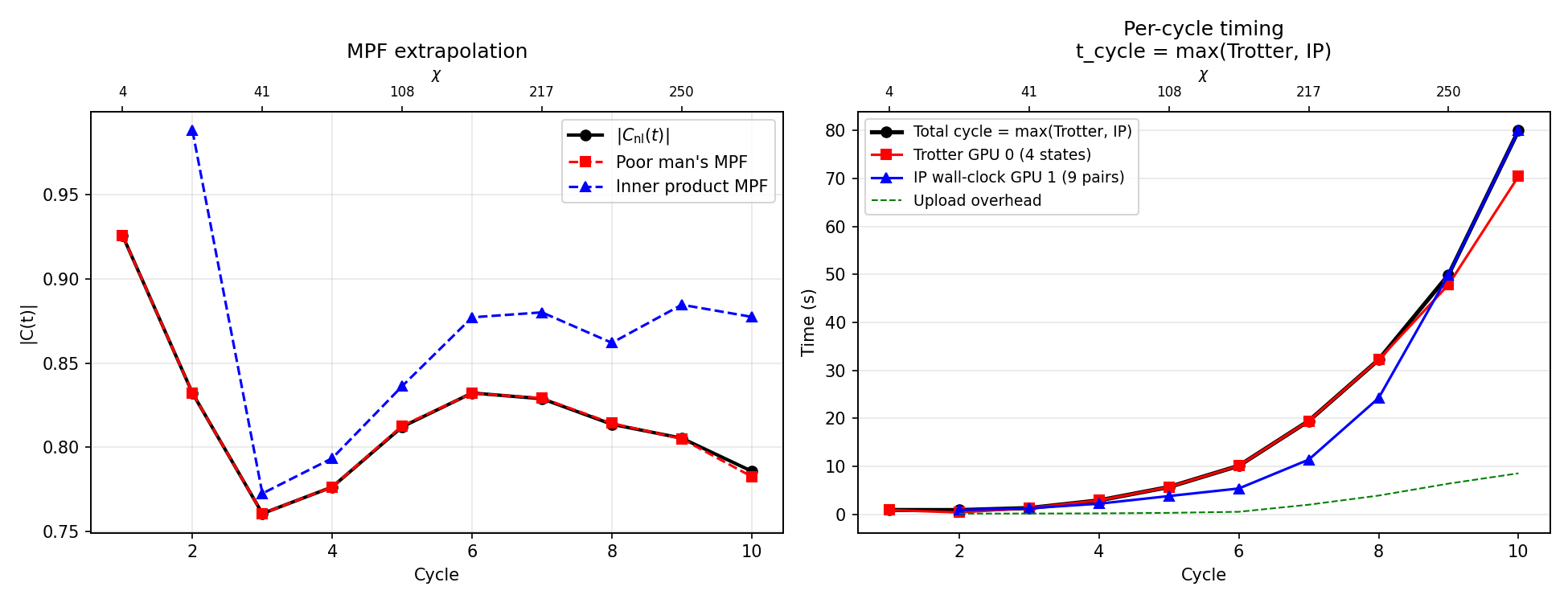}
    \caption{Per-cycle timing at $\chi_{\max}=250$, 10 cycles.
    \emph{Left}: MPF extrapolation — noiseless $C_{\rm nl}$ (black),
    poor man's MPF (red), inner product MPF (blue).
    \emph{Right}: wall-clock breakdown.
    Black: total cycle time $t_{\rm cycle} = \max(t_{\rm Trotter}, t_{\rm IP})$.
    Red: pure Trotter time (4 states, GPU~0).
    Blue: IP wall-clock (9 pairs, parallel streams, GPU~1).
    Green dashed: P2P upload overhead.
    Crossover at cycle 2--3: IP bottleneck early ($\chi\lesssim 41$),
    Trotter dominates thereafter.}
    \label{fig:ip_timing}
\end{figure}

\paragraph{Two-GPU pipeline and timing.}
The 9 inner product pairs ($A_{ij}$ and $B_i$) are independent and
launched as concurrent streams on GPU~1, while GPU~0 runs the four-state
Trotter step.
The total cycle time is $t_{\rm cycle} = \max(t_{\rm Trotter},\, t_{\rm IP})$
(Fig.~\ref{fig:ip_timing}).
Table~\ref{tab:timing} shows the measured breakdown:
Trotter dominates from cycle 3 onward, so inner products add zero wall-clock
overhead for the remainder of the simulation.
At $\chi=250$ (cycle 10): $t_{\rm Trotter}\approx 80$\,s,
$t_{\rm IP}\approx 11$\,s — inner products are $7\times$ faster than Trotter.
Compared to the Julia reference at $350$\,s per cycle,
CppSim achieves a $4\times$ improvement at the same bond dimension.

\begin{table}[h]
\centering
\small
\begin{tabular}{rrrrr}
\toprule
Cycle & $\chi$ & $t_{\rm cycle}$\,(s) & $t_{\rm Trotter}$\,(s) & $t_{\rm IP}$\,(s) \\
\midrule
2  & 18  & 0.89  & 0.39  & 0.89 \\
3  & 41  & 1.31  & 1.31  & 1.26 \\
5  & 108 & 5.69  & 5.69  & 3.81 \\
7  & 214 & 18.44 & 18.44 & 5.39 \\
10 & 250 & 79.68 & 70.48 & 11.00 \\
\bottomrule
\end{tabular}
\caption{Per-cycle wall-clock times at $\chi_{\max}=250$, 10 cycles.
$t_{\rm cycle} = \max(t_{\rm Trotter}, t_{\rm IP})$.
$t_{\rm Trotter}$: GPU~0 Trotter time (4 states sequential).
$t_{\rm IP}$: GPU~1 wall-clock for all 9 inner products (parallel streams).
Trotter dominates from cycle 3 — inner products add zero overhead.}
\label{tab:timing}
\end{table}

\paragraph{Validation of inner product MPF.}
The inner product MPF (Fig.~\ref{fig:ip_timing}, left) exhibits a
\emph{consistent systematic bias}: the extrapolated curve follows the
correct oscillation shape and recurrence structure of $C_{\rm nl}(t)$,
displaced by a roughly constant amplitude offset.
This behaviour distinguishes a correct implementation with a
gauge-dependent approximation error from a broken one — random scatter
would indicate algorithmic failure.
The double-layer BP on the loopy heavy-hex graph approximates the
exact contraction, and the QR gauge introduces a reproducible bias in
the Bethe partition function.
The consistent bias confirms that the $O(\chi^3 D)$ GEMM-based
inner product pipeline is correct; improving accuracy requires
a better gauge (e.g.\ symmetric gauge) or loop corrections,
not algorithmic changes.

\paragraph{Rank-1 message approximation.}
The double-layer message $M_k[i_{a,k}, i_{b,k}]$ couples bond $k$ of state
$|\psi_a\rangle$ to bond $k$ of state $|\psi_b\rangle$.
Note that the bond indices $i_{a,k}$ and $i_{b,k}$ belong to \emph{different}
states — the Kronecker product is across layers, not across bonds.
If $M_k$ is approximated as rank-1:
\begin{equation}
    M_k[i_{a,k}, i_{b,k}] \;\approx\; u_k[i_{a,k}]\cdot v_k[i_{b,k}]
    \label{eq:rank1}
\end{equation}
the two layers decouple completely.
Each contraction in~(\ref{eq:step1}) separates into independent single-layer
operations — $O(\chi^2 D)$ each — identical in structure to the
autocorrelation computation (which is the special case $\chi_b=1$,
$v_k \equiv 1$).
On a tree this rank-1 approximation is exact when messages have converged.
On the loopy heavy-hex graph it is an additional approximation on top of BP,
but the error is bounded by the same loop-correction terms that make plain BP
approximate.
The rank-1 vectors $u_k, v_k$ are updated each sweep via the leading singular
vectors of the full message matrix~(\ref{eq:rank1}), extracted by power iteration
at $O(\chi^2)$ cost.
This approximation reduces the inner product cost to $O(\chi^2 D)$ per message
update — the same order as autocorrelation — and is the natural next step
toward a production GPU implementation.

\paragraph{Symmetric gauge: diagonal messages.}
The strongest simplification arises in the \emph{symmetric gauge}, where
the converged single-layer BP messages are absorbed into the site tensors
as $\psi^{\sqrt{}}_a \leftarrow \sqrt{M_e}\,\psi_a$ for each incident bond.
After this absorption the effective single-layer messages are identity,
and the warm-start~(\ref{eq:msg_init}) gives $M_e^{(0)} = I$ — a diagonal
matrix.
Assuming BP converges back to diagonal matrices (which is forced to be
true on a tree, and is an excellent approximation on the heavy-hex graph
when the gauge is well-chosen), the message product in the vertex factor
$a_v$ is non-zero only when $i_{a,k} = i_{b,k}$ for every bond $k\in\partial v$.
The double sum over $(\mathbf{i}_a, \mathbf{i}_b)$ collapses to a single sum:
\begin{equation}
    a_v \;\approx\; \sum_{\mathbf{i},\,d}
    \psi_a^{\sqrt{}}[v,\mathbf{i},d]\cdot\psi_b^{\sqrt{}}[v,\mathbf{i},d],
    \label{eq:sqrt_av}
\end{equation}
which costs $O(\chi^{\deg} D)$ — identical to the autocorrelation evaluation.
No BP sweeps are required and the $\chi\times\chi$ message matrices never
appear explicitly.
Implementing this gauge change on GPU~1 before the Bethe PF evaluation
would reduce the inner product cost from $O(\chi^3 D)$ to $O(\chi D)$,
making the inner product pipeline faster than Trotter at all bond dimensions.

\begin{figure}[ht]
    \centering
    \includegraphics[width=\columnwidth]{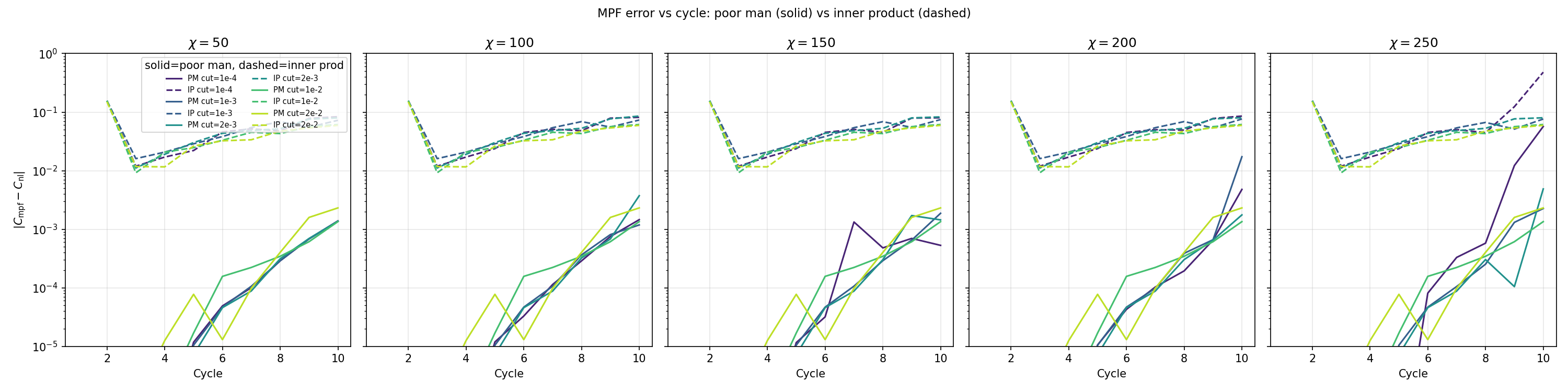}
    \caption{MPF error $|C_{\rm mpf}(t) - C_{\rm nl}(t)|$ vs cycle for
    $\chi_{\max} \in \{50,100,150,200,250\}$ and
    cutoff $\in \{2\times10^{-2}, 10^{-2}, 2\times10^{-3}, 10^{-3}, 10^{-4}\}$.
    Solid lines: poor man's MPF. Dashed lines: inner product MPF.
    \emph{Poor man's MPF} shows significant variation across chi and cutoff
    — it genuinely improves with tighter approximation.
    \emph{Inner product MPF} curves cluster tightly regardless of chi or cutoff
    — the error is gauge-limited, not compute-limited.
    Both methods show error growth at $\chi=250$ cycle~10 where chi saturates
    and truncation error accumulates — a physics limit, not an MPF artifact.}
    \label{fig:ip_mpf}
\end{figure}

\begin{figure}[ht]
    \centering
    \includegraphics[width=0.85\columnwidth]{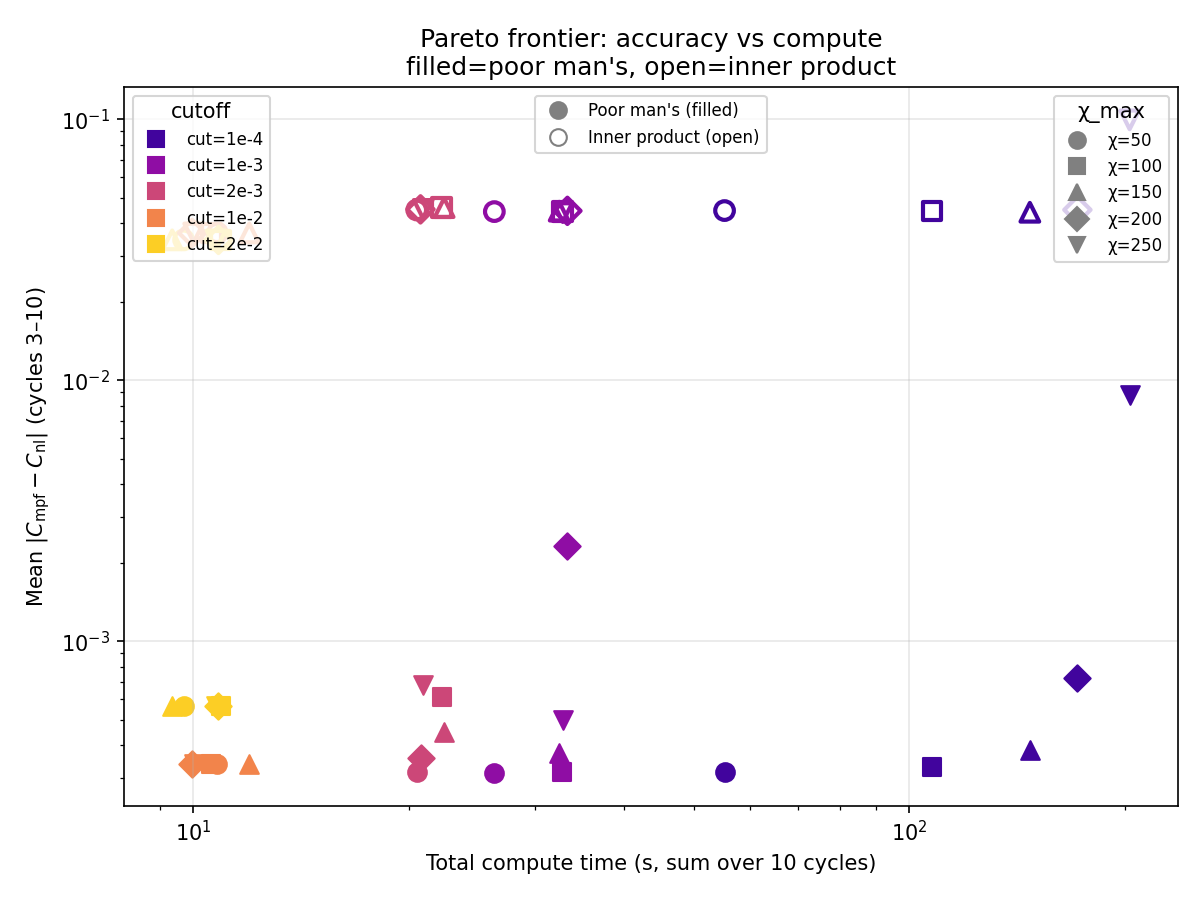}
    \caption{Pareto frontier: mean error (cycles 3--10) vs total compute time
    for all 25 (chi, cutoff) combinations.
    Filled markers: poor man's MPF. Open markers: inner product MPF.
    Shape encodes $\chi_{\max}$; color encodes cutoff.
    \emph{The Pareto frontier is entirely dominated by poor man's MPF}
    — no inner product point achieves competitive accuracy at any compute budget.
    Poor man's MPF achieves mean error $\sim 3\times10^{-4}$ across the full
    compute range; inner product MPF stays at $\sim 5\times10^{-2}$,
    two orders of magnitude worse.
    The inner product points are vertically flat (insensitive to chi/cutoff),
    confirming that the error is set by the QR gauge approximation bias,
    not by approximation quality.}
    \label{fig:ip_pareto}
\end{figure}

The polynomial extrapolation is a Taylor approximation in $\gamma$, valid when
the noise perturbation is small. As cycles accumulate, the noise effect grows
and higher-order terms become significant, causing the polynomial fit to diverge.
The poor man's MPF (correlation matrix) exploits the full temporal structure of
the signal and outperforms the polynomial by 2--3 orders of magnitude up to
cycle 12 (Fig.~\ref{fig:mpf}).
The full 10-cycle gamma-MPF computation at $\chi=70$ completes in under 5\,s on a 32\,GB GPU.


\section{Conclusion}
\label{sec:conclusion}

Our design maintains the original idea of being topology-agnostic: gate
and belief propagation kernels require no modification; only a new
grid class and Hamiltonian gate matrix are needed for the new geometry.
To showcase different physics, we incorporate a fully parametric hardware
noise model: each bond carries an independent $16\times16$ Pauli transfer
matrix per color class, loaded from device characterization data and
reduced into a single matrix--vector product per bond — at no additional
runtime cost over the noiseless case. This approach captures bond
interactions without assumptions on computation order, and is
mathematically equivalent to sequential gate application.

Our goal is to provide a framework where physics can be expressed
directly for GPU, maintaining the intuitive formulation of the
Julia/AMDGPU.jl reference \cite{rudolph2025simulating} while
explicitly exploiting parallelism and performance.
For quantum error mitigation via the gamma multi-product formula
\cite{mpf-ref}, CppSim completes 20 Trotter cycles in a fraction of
the time of current implementations, enabling fast and systematic
exploration of the parameter space.

Future work will extend to larger graphs, implement the full quantum
inner product coefficient matrix via GPU double-layer belief
propagation, and explore the joint epsilon--gamma MPF for simultaneous
Trotter and noise extrapolation.
We are also interested in extending the lightcone prediction and
modeling to the noisy setting using the same methodology.

\bibliographystyle{unsrt}
\bibliography{heavyhex}

@misc{dalberto2026cppsim,
      title={GPU-First Heisenberg-Picture Tensor Network Dynamics for the 2D Transverse-Field Ising Model},
      author={Paolo D'Alberto},
      year={2026},
      eprint={2606.30985},
      archivePrefix={arXiv},
      primaryClass={cs.MS},
      url={https://arxiv.org/abs/2606.30985},
}

@misc{rudolph2025simulating,
  title         = {Simulating and Sampling from Quantum Circuits with 2D Tensor Networks},
  author        = {M. S. Rudolph and J. Tindall},
  year          = {2025},
  eprint        = {2507.11424},
  archivePrefix = {arXiv},
  primaryClass  = {quant-ph},
  url           = {https://arxiv.org/abs/2507.11424}
}

@article{tindall2023gauging,
  title     = {Gauging tensor networks with belief propagation},
  author    = {J. Tindall and M. Fishman},
  journal   = {SciPost Physics},
  volume    = {15},
  pages     = {222},
  year      = {2023},
  doi       = {10.21468/SciPostPhys.15.6.222}
}

@article{tindall2024eagle,
  title     = {Efficient Tensor Network Simulation of {IBM}'s Eagle Kicked Ising Experiment},
  author    = {J. Tindall and M. Fishman and E. M. Stoudenmire and D. Sels},
  journal   = {PRX Quantum},
  volume    = {5},
  pages     = {010308},
  year      = {2024},
  doi       = {10.1103/PRXQuantum.5.010308}
}

@article{evenbly2026loop,
  title     = {Loop Series Expansions for Tensor Networks},
  author    = {G. Evenbly and N. Pancotti and A. Milsted and J. Gray and G. K.-L. Chan},
  journal   = {Physical Review Research},
  volume    = {8},
  pages     = {013245},
  year      = {2026},
  doi       = {10.1103/PhysRevResearch.8.013245}
}

@misc{tindall2025disordered,
  title         = {Dynamics of disordered quantum systems with two- and three-dimensional tensor networks},
  author        = {J. Tindall and A. Mello and M. Fishman and M. Stoudenmire and D. Sels},
  year          = {2025},
  eprint        = {2503.05693},
  archivePrefix = {arXiv},
  primaryClass  = {quant-ph},
  url           = {https://arxiv.org/abs/2503.05693}
}

@article{heavyhex-topology,
  title     = {Topological and Subsystem Codes on Low-Degree Graphs with Flag Qubits},
  author    = {C. Chamberland and G. Zhu and T. J. Yoder and J. B. Hertzberg and A. W. Cross},
  journal   = {PRX Quantum},
  volume    = {2},
  pages     = {030345},
  year      = {2021},
  doi       = {10.1103/PRXQuantum.2.030345}
}

@misc{mpf-ref,
  title         = {Well-conditioned multi-product formulas for hardware-friendly {Hamiltonian} simulation},
  author        = {G. H. Low and V. Kliuchnikov and N. Wiebe},
  year          = {2019},
  eprint        = {1907.11679},
  archivePrefix = {arXiv},
  primaryClass  = {quant-ph},
  url           = {https://arxiv.org/abs/1907.11679}
}

\end{document}